# Impulse Measurement Methods for Pulsed Laser Ablation Propulsion

Ishaan Mishra[1], Scott Kirkpatrick[2]

[1] mishrai@rose-hulman.edu, Physics and Optical Engineering, Rose-Hulman Institute of Technology
[2] kirkpat2@rose-hulman.edu, Physics and Optical Engineering, Rose-Hulman Institute of Technology

**Abstract**

Pulsed laser ablation propulsion has the potential to revolutionize space exploration by eliminating the requirement of a spacecraft to carry its propellant and power source as the high-power laser is situated off-board. More experimentation needs to be done to optimize this propulsion system and understand the mechanisms of thrust generation. There are many methods used to calculate the impulse imparted in pulsed laser ablation experiments. In this paper, key performance parameters are derived for some of the impulse measurement methods used in ablation propulsion experiments. Regimes discussed include the torsional pendulum system, simple pendulum system, and solid and liquid microspheres.

## Introduction

The history of research in laser ablation propulsion spans the last 40 years and is a technology that enables spacecraft to not require a power or a propellant source to be carried on board [1]. Traditional propulsion methods rely on energizing propellant it carries through chemical, nuclear, or electromagnetic sources of energy. The former two require a large propellant mass to be carried by the spacecraft, while the latter requires a low propellant mass but a large power source. Laser-driven propulsion eliminates the requirement of an on-board power system, while the target propellant can be sourced from the nearby environment [2]. A high-intensity laser strikes the target surface which causes surface particles to be vaporized, entering a plasma phase. The expansion of this ablated material provides thrust to the spacecraft. While propellant mass is still required, the energy source required to activate the propellant to produce thrust is not carried by the spacecraft. Typically, pulse width ranges between nanoseconds to femtoseconds. Other applications of laser ablation include space debris clearing [3], inertial fusion energy [4], as well as air-breathing propulsion [5].

Various measurement issues exist in pulsed laser ablation propulsion [6]. In this paper, a summary of the major impulse measurement methods in the field are described. This paper serves to be a resource for future experimentation in the field of pulsed laser ablation propulsion.

# Theory of Pulsed Laser Ablation Propulsion

## I. Performance Parameters

First, key performance parameters associated with laser ablation propulsion are discussed, which refers heavily to the review on laser ablation propulsion by Phipps et. al. in 2010 [1]. The momentum coupling coefficient $C_m$ is a performance parameter that is the total imparted momentum per input unit laser energy defined as the impulse density of the laser ($\sigma$) divided by the laser pulse fluence ($\Phi$).

$$C_m = \frac{\sigma}{I} = \frac{\mu v_E}{\phi} = \frac{p}{I} \qquad (1)$$

where $I$ is the intensity of the laser, and $p$ is the pressure in the case of a continuous wave system [1], $\mu$ is the areal mass density, and $v_E$ is the exhaust velocity. The specific ablation energy $Q^*$ is defined by dividing the laser fluence by the target areal mass density.

$$Q^* = \frac{\phi}{\mu} \qquad (2)$$

The ablation efficiency is defined as

$$\eta_{AB} = \frac{\mu \psi v_E^2}{2\phi} = \frac{\psi C_m v_E}{2} \qquad (3)$$

where $\psi$ is defined in [7] as,

$$\psi = \frac{<v_x^2>}{(<v_x>)^2} = \left\{ \frac{u^2 + \frac{kT}{m_E}}{u^2} \right\} \qquad (4)$$

Thus, the momentum coupling coefficient and specific impulse ($I_{sp}$) satisfy the relation

$$C_m I_{sp} = \frac{2\eta_{AB}}{\psi g_0} \qquad (5)$$

where $g_0$ is the acceleration due to gravity on the surface of the Earth and

$$I_{sp} = \frac{v_E}{g_0} \qquad (6)$$

When a high-intensity laser strikes the surface of a substance, the ablated material is usually a combination of its vapor regime, as well as its plasma regime. To study the plume characteristics and shockwave, we must understand both the plasma and vapor treatment of the ablated material. In most applied cases, the characteristics of the ablated propellant are a combination of the plasma and vapor regime.

A. Plasma Regime

For ablated material with sufficient energy that the ionization fraction approaches 1, the momentum coupling coefficient and specific impulse for the plasma-dominated regime are given below [8]

$$C_{m,p} = 5.83 \frac{\psi^{\frac{9}{16}}}{A^{\frac{1}{8}}(I\lambda\sqrt{\tau})^{\frac{1}{4}}} \, dyn/W \tag{7}$$

$$I_{sp,p} = 1400 \frac{A^{\frac{1}{8}}}{\psi^{\frac{9}{16}}} (I\lambda\sqrt{\tau})^{\frac{1}{4}} \, s \tag{8}$$

for

$$\psi = \frac{A}{2[Z^2(Z+1)]^{\frac{1}{3}}} \tag{9}$$

where A is the atomic mass number and Z is the average ionization state in the plasma plume, derived using the Saha Equation

$$\frac{n_e n_j}{n_{j-1}} = \frac{2u_j}{u_{j-1}} \left(\frac{2\pi A m_p k T_e}{h^2}\right)^{\frac{3}{2}} \exp\left(-\frac{W_{j,j-1}}{kT_e}\right) \tag{10}$$

for

$$Z = \frac{1}{n_i} \sum_{j=1}^{j\,max} (j n_j) \geq 1 \tag{11}$$

and

$$\sum_{j=1}^{j\,max} (n_j) = n_i \tag{12}$$

where $W_{j,j-1}$ is the ionization energy difference between the (j-1)th and jth ionization states of the material, $m_p$ is the proton mass, $n_o$ is the neutral vapor density, $u_j$ is the quantum mechanical partition function of the jth state, and $n_j$ is the number density of each of the ionized states.

B. Vapor Regime

For simplicity, the vapor is assumed to be an elemental material. Many propellants are polymers, for which a treatment is given in [1].

$$C_{m,v} = \frac{p}{I} \tag{13}$$

$$I_{sp,v} = \frac{v}{g_o} \tag{14}$$

C. Combined Regime

In PLAP, the performance parameters are a combination of the vapor and plasma regime. This is because all ablated material does not become a plasma. Thus, for ablated material where $\eta_i$ is the ionization fraction, the momentum coupling coefficient and specific impulse are simply given by

$$C_m = (1 - \eta_i) C_{m,v} + \eta_i C_{m,p} \tag{15}$$

$$I_{sp} = (1 - \eta_i) I_{sp,v} + \eta_i I_{sp,p} \tag{16}$$

## II. Thrust Measurement Mechanics

In pulsed laser propulsion, a high-intensity laser incident on a surface has different thrust generation mechanism based on the medium the target surface is in-

- In a vacuum causes material surface to vaporize, forming a plasma. Momentum transfer occurs due to the evaporation and plasma expansion at the surface material [2]. A simplified diagram of the plasma expansion is shown in figure 1. The plasma is generated through multi-photon ionization and avalanche ionization [9].
- In a fluid medium (like air, or a liquid), if the laser energy density is greater than the fluid's ionization threshold momentum transfer also occurs by the shock wave expansion in the medium, which is why laser ablative propulsion has a higher efficiency.

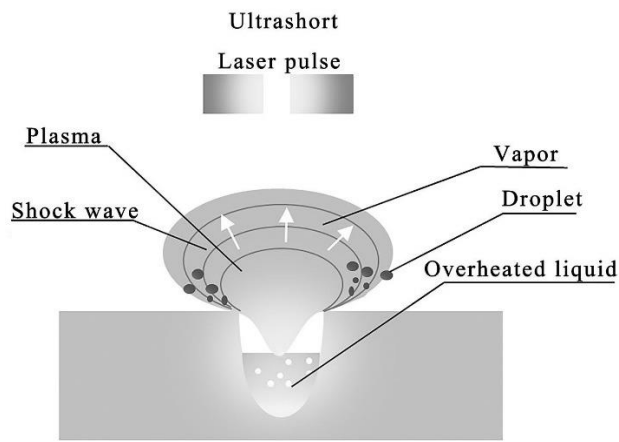

Figure 1: Ultrashort laser pulse beam-matter interaction [2]

## Impulse Measurement Methods

In this paper, the focus for measuring motion of ablated targets is on impulse measurement, not thrust measurement. The reason is that most of the thrust is imparted due to the motion of the ablated material in vapor/plasma state and is miniscule and varies with time. While the pulse duration is known, it is unclear what the time scale of the plasma-material interaction and shockwave expansion phenomena are, which is significant since these are the dominant thrust-generation mechanisms. Since we don't know the duration over which the thrust is imparted, it is challenging to record the thrust. Thus, impulse measurement is more practical to characterize pulsed laser ablation propulsion. For each of the methods discussed below, equations for the specific impulse and momentum coupling coefficient are highlighted.

I. Torsion Pendulum

In this system, a balancing arm is suspended by a thin wire, and the laser target is placed at one end of the arm. Upon being irradiated by the laser pulse(s), ablated material is ejected which creates a torque on the system. A diagram of this system is shown in Figure 2.

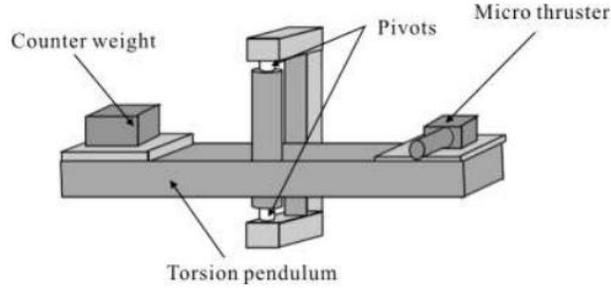

Figure 2: Sample torsional pendulum setup, where the ablation target is placed in place of the micro thruster [10]

In vacuum, the momentum in the ablation process is calculated with the following relation

$$p = \frac{\theta_{max}\sqrt{J_T I}}{r} \quad (17)$$

Where $\theta_{max}$ is the maximum displaced angle recorded, $J_T$ is the torsion constant, of the wire, $I$ is the moment of inertia, and $r$ is the distance of the ablation region from the center of the pendulum arm. The momentum coupling coefficient here is derived from the relation in Equation 1-

$$C_m = \frac{p}{E} \quad (18)$$

where E is the energy of the laser pulse(s) [11, 12]. The specific impulse is calculated with the equation

$$I_{sp} = \frac{p}{\Delta m g_0} \quad (19)$$

where $\Delta m$ is the mass of ablated material. This can be calculated by measuring the ablated volume with a microscope. $g_0$ is the standard acceleration of the Earth's gravity [13]. For a more detailed analysis of the thrust, one can analyze the angular displacement of the torsional pendulum with respect to time. An underdamped oscillating torsion pendulum system can be described by the equation below

$$I\ddot{\theta}(t) + B\dot{\theta}(t) + K\theta(t) = F(t) \quad (20)$$

where $\theta(t)$ is the angular displacement of the arm of the torsional pendulum. This angle is measured by dividing the recorded linear displacement of the arm and dividing it by the rotating radius (small angle approximation) [14]. A note is that with smaller angular displacement, the error due to noise is very large. The angular displacement can be measured with a laser interferometer [15] or a magnetic encoder. The values of $I$, $B$, and $K$ can be calculated by applying a known impulse and characterizing the motion of the pendulum. A method to do so is described here –

1) A known impulse is applied at the end of the torsional pendulum.
2) The motion of the pendulum is recorded, and a best fit equation of motion is described with the to match the equation (assuming underdamped pendulum system)

$$\theta(t) = u(t-a) \cdot e^{A(a-t)} [C_1 \sin(t-a) + C_2 \cos(t-a)] \qquad (21)$$

where $a$ is the time at which the impulse is imparted, $C_1$, $C_2$, and $A$ are unknown values obtained in the best fit.

3) Taking the Laplace transform of the transfer equation of Equation 21 and equating it to the Laplace transform of the best fit equation of motion obtained in step 2, the values of $I$, $B$, and $K$ are found [16].

Once these values are known, the thrust $F(t)$ can be described.

## II. Simple Pendulum

In a simple pendulum system, an ultrashort pulse strikes the ablative target which is suspended by a pendulum of length $L$. Upon ablation, the target swings, from which the momentum coupling coefficient can be calculated with the equation derived from theory established in [17]

$$C_m = \frac{m g_0 r T}{2\pi E L} \sqrt{2(1 - \cos(\theta_{max}))} \qquad (22)$$

Where m is the mass of the pendulum, g is the acceleration due to gravity, r is the distance of the suspension point to the center of mass, T is the time period, E is the energy of the laser pulse(s), and $\theta_{max}$ is the maximum displacement angle of the pendulum. The specific impulse is derived to be

$$I_{sp} = \frac{m}{\Delta m} \sqrt{\frac{2h}{g_0}} \qquad (23)$$

Error in calculating the ablated mass $\Delta m$ results in high errors in the values of the specific impulse. Data of the motion of the pendulum can be collected with a camera that records the setup, as shown below in Figure 3.

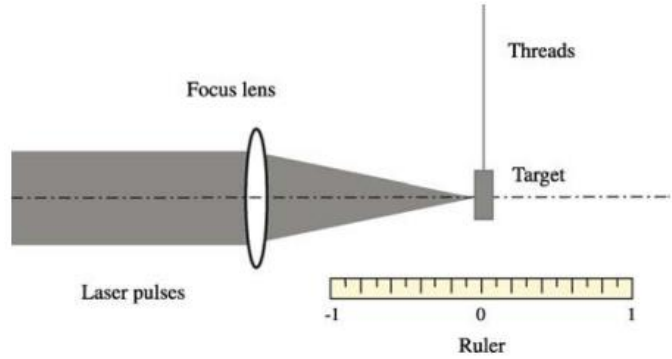

Figure 3: Simple pendulum system [18]

### III. Solid microspheres

There are several methods to use solid microspheres to understand impulse of laser ablation. The first method discussed here is that of projectile motion of an ablated target. A microsphere is placed at the edge of a known height h. The laser pulse is directed to the center of the microsphere. This results in the ablative process that causes the microsphere to displace, undergoing projectile motion. The horizontal distance it covers is measured, and the momentum of the microsphere can be calculated. The setup is shown below in Figure 4.

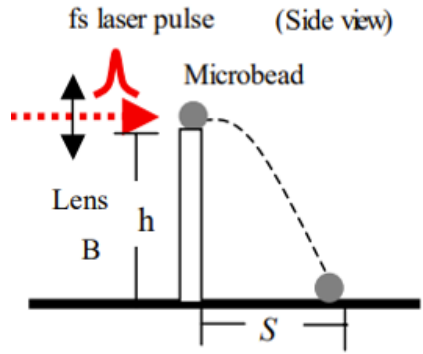

Figure 4: Expected projectile motion of a solid microsphere ablation target upon being struck by a laser [19]

The momentum coupling coefficient and specific impulse are given by the equations below

$$C_m = \frac{mS}{E}\sqrt{\frac{g_0}{2h}} \tag{23}$$

$$I_{sp} = \frac{m}{\Delta m}\frac{S}{\sqrt{2g_0}} \tag{24}$$

Where $m$ is the mass of the microbead, $\Delta m$ is the ablated mass, $S$ is the horizontal displacement, $h$ is the vertical displacement, $g_0$ is the acceleration due to gravity, and $E$ is the energy of a laser pulse. A time-resolved imaging system as described in [20] can be used to observe the motion, and to record the location the microbead makes contact with the base, before it drifts or rolls further.

The second method is one used by Yu et. al. in [21] to explore the interaction of nanosecond laser pulse on microbeads, whose linear motion on a substrate are observed. While this method is useful in observing the plasma characteristics, due to the movement mechanism of the microspheres, information of impulse, and coupling coefficient are hard to ascertain. This experiment proved that laser-ablated propulsion is mainly dominated by shock wave ejection mechanism. The motion of the microspheres is a combination of rolling and sliding action, but the exact contribution of each is "almost impossible" to calculate [22].

Demos et. al. in [23] describe another experiment, wherein stainless-steel microspheres of diameter 27 micrometer where placed on a silicon dioxide substrate whose surface is parallel to gravity. The laser shown on the microsphere target results in confined and subsequently semi-confined plasma expansion that results in the propulsion of the microsphere. The motion of the microsphere is recorded with a time-resolved shadowgraphic microscope system. Key performance parameters like momentum coupling coefficient, velocity of the microbeads, etc. are obtained from the data.

## IV. Liquid microspheres

In recent years, pulsed laser ablation propulsion experiments have also been conducted on liquid microspheres of various metallic and non-metallic materials [24, 25, 26]. One or more laser pulses are irradiated on microdroplets that are moving at a known speed. The plasma generation, expansion, and effect on the droplet shape and motion is then observed with time-resolved imaging. Since the volume and density of the substances are known, calculations of the velocity, momentum, momentum coupling coefficient, and efficiency can be made, as well as data on plasma plume formation. The discussion of liquid microspheres is limited as it is not directly related to space propulsion applications, which is the focus of this paper.

## V. Other Methods

Two additional methods are briefly mentioned here.

- The time-of-flight method helps in understanding the velocity distribution of ablated ions. An example of the TOF method can be found in the work of Mengqi and Grigoropoulos [27] where ablation mechanisms and crater formation are analyzed using this method. Phipps et al. [28] analyze data using the TOF method as well as a torsional pendulum. The TOF method in this experiment helped in analyzing the ion velocity peaks, which aid in characterizing the ablated plasma.
- Piezoelectric pressure sensors have a very high time resolution unlike the torsional pendulum and gravitational pendulum system and have been used in experiments in [29] and [30]. Thus, for experiments which require high temporal resolution, commercially available piezoelectric sensors are preferred.

## Conclusion

In this paper, commonly used methods for impulse measurement methods used in pulsed laser ablation propulsion experiments have been described. Torsion and simple pendulums, solid and liquid microspheres, time-of-flight method, and piezoelectric sensors have all been discussed. Based on the requirements of experiments, one of the above impulse measurement methods can be used, or a novel method can be deployed.

## References

[1]: Phipps, Claude, et al. "A review of laser ablation propulsion." *AIP Conference Proceedings*. Vol. 1278. No. 1. American Institute of Physics, 2010.

[2]: Yu, Haichao, et al. "Brief review on pulse laser propulsion." *Optics & Laser Technology* 100 (2018): 57-74.

[3]: Phipps, C. R., et al. "ORION: Clearing near-Earth space debris using a 20-kW, 530-nm, Earth-based, repetitively pulsed laser." *Laser and Particle Beams* 14.1 (1996): 1-44.

[4]: Nakai, S., and K. Mima. "Laser driven inertial fusion energy: present and prospective." *Reports on Progress in Physics* 67.3 (2004): 321.

[5]: Myrabo, Leik. "World record flights of beam-riding rocket lightcraft-Demonstration of" disruptive" propulsion technology." *37th Joint Propulsion Conference and Exhibit*. (2001).

[6]: Sinko, John E., et al. "Measurement issues in pulsed laser propulsion." *AIP Conference Proceedings*. Vol. 1230. No. 1. American Institute of Physics, (2010).

[7]: Phipps, Claude R., James P. Reilly, and Jonathan W. Campbell. "Optimum parameters for laser launching objects into low Earth orbit." *Laser and particle beams* 18.4 (2000): 661-695.

[8]: Phipps Jr, C. R., et al. "Impulse coupling to targets in vacuum by KrF, HF, and CO2 single-pulse lasers." *Journal of Applied Physics* 64.3 (1988): 1083-1096.

[9]: Phipps, Claude R., et al. "Micropropulsion using laser ablation." *Applied Physics A* 79.4 (2004): 1385-1389.

[10]: Xing, Jin, et al. "Design method of micro-impulse measuring system based on pivots for pulsed micro-thruster." *红外与激光工程* 48.S1 (2019): 97-103.

[11]: Zhu, Xiao-nong, and Nan Zhang. "Investigation of ultrashort pulse laser propulsion using time-resolved shadowgraphy and torsion pendulum." *International Symposium on Photoelectronic Detection and Imaging 2009: Laser Sensing and Imaging*. Vol. 7382. SPIE (2009).

[12]: Li, Jing, and Zhiping Tang. "Laser micro-impulse torsion pendulum." *Chinese Optics Letters* 3.2 (2005): 76-79.

[13]: Cai, Jian, Xiaojun Hu, and Zhiping Tang. "An experimental and conceptual investigation of laser micro propulsion." *AIP Conference Proceedings*. Vol. 830. No. 1. American Institute of Physics, 2006.

[14]: D'Souza, Brian C., and Andrew D. Ketsdever. "Investigation of time-dependent forces on a nano-Newton-second impulse balance." *Review of Scientific Instruments* 76.1 (2005): 015105.

[15]: Li, Yan-Chao, and Chun-Hui Wang. "A method of measuring micro-impulse with torsion pendulum based on multi-beam laser heterodyne." *Chinese Physics B* 21.2 (2012): 020701.

[16]: Noonburg, Virginia W. Differential Equations: From Calculus to Dynamical Systems. Vol. 43. American Mathematical Soc., (2020).

[17]: Saeed, Humaima, et al. "Quantitative Measurements of Ablative Laser Propulsion Parameters of Metal Foils Using Pulsed Nd: YAG Laser." *Arabian Journal for Science and Engineering* (2021): 1-7.

[18]: Zheng, Z. Y., et al. "Characteristic investigation of ablative laser propulsion driven by nanosecond laser pulses." *Applied Physics A* 83.2 (2006): 329-332.

[19]: Zhang, Nan, You-Bo Zhao, and Xiao-Nong Zhu. "Light propulsion of microbeads with femtosecond laser pulses." *Optics express* 12.15 (2004): 3590-3598.

[20]: Raman, Rajesh N., Raluca A. Negres, and Stavros G. Demos. "Time-resolved microscope system to image material response following localized laser energy deposition: exit surface damage in fused silica as a case example." *Optical Engineering* 50.1 (2011): 013602.

[21]: Yu, Haichao., et al. "Dynamic testing of nanosecond laser pulse induced plasma shock wave propulsion for microsphere." *Applied Physics A* 126.1 (2020): 1-8.

[22]: Hooper, Thomas, and Cetin Cetinkaya. "Efficiency studies of particle removal with pulsed-laser induced plasma." Journal of adhesion science and technology 17.6 (2003): 763-776.

[23]: Demos, Stavros G., et al. "Mechanisms governing the interaction of metallic particles with nanosecond laser pulses." *Optics Express* 24.7 (2016): 7792-7815.

[24]: Kurilovich, DmitSry, et al. "Plasma propulsion of a metallic microdroplet and its deformation upon laser impact." *Physical review applied* 6.1 (2016): 014018.

[25]: Klein, Alexander L., et al. "Drop shaping by laser-pulse impact." *Physical review applied* 3.4 (2015): 044018.

[26]: Chen, Ziqi, et al. "Investigation of Nd: YAG laser produced tin droplet plasma expansion." *Laser Physics Letters* 13.5 (2016): 056002.

[27]: Ye, Mengqi, and Costas P. Grigoropoulos. "Time-of-flight and emission spectroscopy study of femtosecond laser ablation of titanium." *Journal of Applied Physics* 89.9 (2001): 5183-5190.

[28]: Phipps, Claude., et al. "Measurements of laser impulse coupling at 130 fs." *High-Power Laser Ablation V*. Vol. 5448. SPIE, 2004.

[29]: Igarashi, Kazuki, et al. "Performance of laser ablation propulsion with a high-repetition rate and high-power laser." *AIP Advances* 12.2 (2022): 025219.

[30]: Pakhomov, Andrew V., et al. "Ablative laser propulsion: specific impulse and thrust derived from force measurements." *AIAA journal* 40.11 (2002): 2305-2311.